\documentclass[twocolumn,preprintnumbers, amsmath,amssymb, bibnotes, nofootinbib]{revtex4}
\usepackage{amsthm}
\usepackage{amsmath}
\usepackage{amssymb}
\usepackage{graphicx}
\usepackage{url}
\usepackage{dsfont}
\usepackage{float}
\usepackage{bm}
\usepackage{ifthen}
\usepackage[usenames,dvipsnames]{color}
\usepackage{mathrsfs}
\usepackage[colorlinks=true,citecolor=blue,urlcolor=black]{hyperref}
\usepackage{float}
\usepackage{afterpage}
\usepackage{subfigure}
\usepackage{times}

\newcommand{\ie}{{i.e.}}

\newcommand{\be}{\begin{equation}}
\newcommand{\ee}{\end{equation}}


\newcommand{\sket}[1]{{\ensuremath{\lvert#1\rangle}}}
\newcommand{\lket}[1]{{\ensuremath{\left\lvert#1\right\rangle}}}
\newcommand{\ket}[1]{\if@display\lket{#1}\else\sket{#1}\fi}
\newcommand{\sbra}[1]{{\ensuremath{\langle#1\rvert}}}
\newcommand{\lbra}[1]{{\ensuremath{\left\langle#1\right\rvert}}}
\newcommand{\bra}[1]{\if@display\lbra{#1}\else\sbra{#1}\fi}
\newcommand{\sbraket}[2]{{\ensuremath{\langle#1\rvert#2\rangle}}}
\newcommand{\lbraket}[2]{{\ensuremath{\left\langle#1\!\left\rvert\vphantom{#1}#2\right.\!\right\rangle}}}
\newcommand{\braket}[2]{\if@display\lbraket{#1}{#2}\else\sbraket{#1}{#2}\fi}

\newcommand{\sketbra}[2]{{\ensuremath{\lvert #1\rangle\!\langle #2\rvert}}}
\newcommand{\lketbra}[2]{{\ensuremath{\left\lvert #1\right\rangle\!\!\left\langle #2\right\rvert}}}
\newcommand{\ketbra}[2]{\if@display\lketbra{#1}{#2}\else\sketbra{#1}{#2}\fi}

\theoremstyle{plain}

\theoremstyle{definition}

\begin{document}
\title{Measurement-device-independent quantum communication with an untrusted source}

\author{Feihu Xu}
 \email{fhxu@mit.edu}
\affiliation{Research Laboratory of Electronics, Massachusetts Institute of Technology, 77 Massachusetts Avenue, Cambridge, Massachusetts 02139, USA}

\date{\today}
\begin{abstract}
Measurement-device-independent quantum key distribution (MDI-QKD) can provide enhanced security, as compared to traditional QKD, and it constitutes an important framework for a quantum network with an untrusted network server. Still, a key assumption in MDI-QKD is that the sources are trusted. We propose here a MDI quantum network with a single untrusted source. We have derived a complete proof of the unconditional security of MDI-QKD with an untrusted source. Using simulations, we have considered various real-life imperfections in its implementation, and the simulation results show that MDI-QKD with an untrusted source provides a key generation rate that is close to the rate of initial MDI-QKD in the asymptotic setting. Our work proves the feasibility of the realization of a quantum network. The network users need only low-cost modulation devices, and they can share both an expensive detector and a complicated laser provided by an untrusted network server.
\end{abstract}
\maketitle

\section{Introduction}
The global quantum network is believed to be the next-generation information-processing platform for speedup computation and a secure means of communication. Among the applications of the quantum network, quantum key distribution
(QKD) is one of the first technology in quantum information science to produce practical applications~\cite{gisin2002quantum,scarani2009security,lo2014secure}. Commercial QKD systems have appeared on the market~\cite{IDQ:company,AnhuiCommercial}, and QKD networks have been developed~\cite{sasaki2011field,frohlich2013quantum,hughes2013network}. Unfortunately, due to real-life imperfections, a crucial problem in current QKD implementations is the discrepancy between its theory and practice~\cite{lo2014secure}. An eavesdropper (Eve) could exploit such imperfections and hack a QKD system. Indeed, the recent demonstrations of various attacks~\cite{Gisin:attack:2006,Xu:phaseremapping:2010, sun2011passive,Yi:timeshift:2008,Lars:nature:2010, Gerhardt:2010, weier2011quantum, jain2011device} on practical QKD systems highlight that the theory-practice discrepancy is a major problem for practical QKD.

Measurement-device-independent quantum key distribution (MDI-QKD)~\cite{Lo:MDIQKD} removes all detector side-channel attacks. This kind of attack is arguably the most important security loophole in conventional QKD implementations~\cite{Yi:timeshift:2008, Lars:nature:2010, Gerhardt:2010, weier2011quantum, jain2011device}. The assumption in MDI-QKD is that the state preparation can be trusted. Unlike security patches~\cite{yuan2011resilience, ferreira2012real, limrandom} and device-independent QKD~\cite{acin2007device}, MDI-QKD can remove all detector loopholes and is also practical for current technology. Hence, MDI-QKD has attracted a lot of scientific attention in both theoretical~\cite{ma2012statistical, ma2012alternative, wang2013three, Feihu:practical, marcos:finite:2013, xu2013long, xu2014protocol, PhysRevLett.114.090501} and experimental~\cite{Tittel:2012:MDI:exp, Liu:2012:MDI:exp, da2012proof, zhiyuan:experiment:2013, tang2014measurement, valivarthi2015measurement} studies. See~\cite{xu2014measurement} for a review of its recent development.


\begin{figure}
\begin{center}
 \includegraphics[scale=0.72,angle=0]{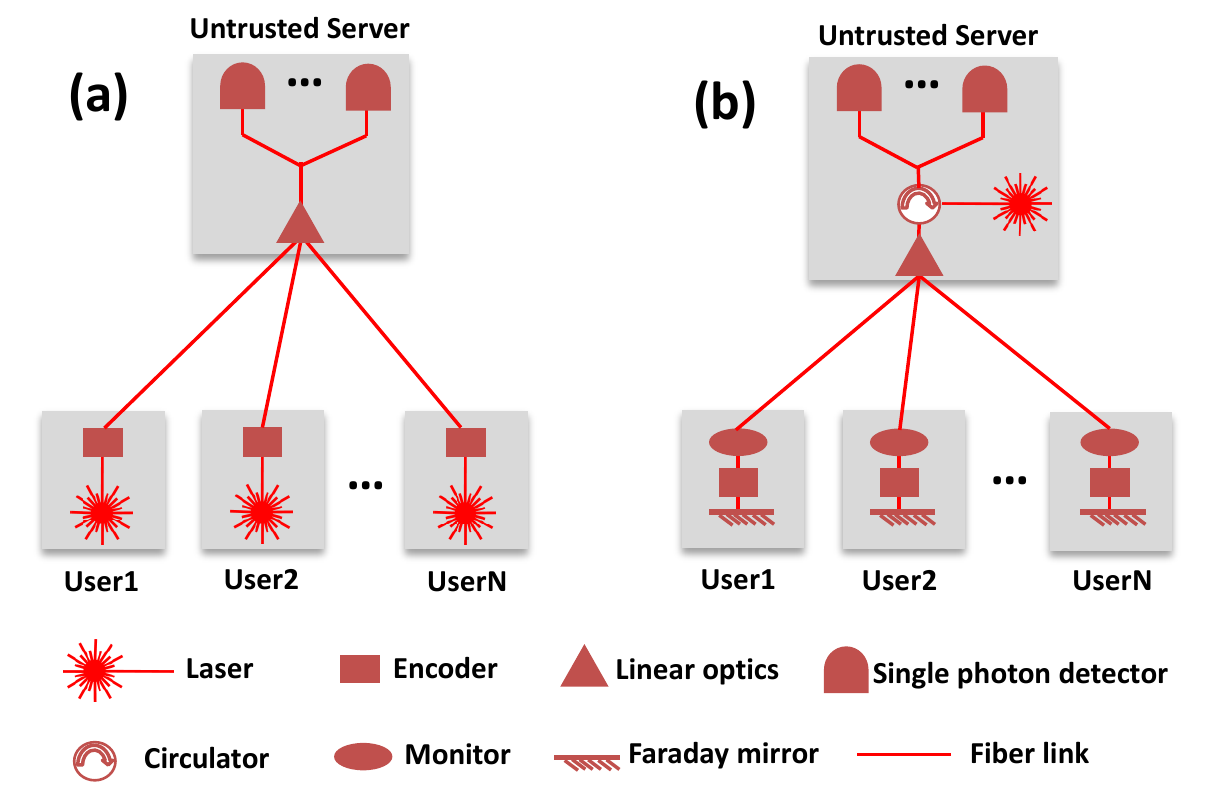}
 \end{center}
 \caption{(Color online) (a) A fiber-based quantum network with N trusted lasers. (b) A quantum network with a single \emph{untrusted} laser source. \label{Fig:network}}
\end{figure}

An important feature of MDI-QKD is that it can be used to build a fiber-based MDI quantum network with a fully \emph{untrusted} network server (see Fig.~\ref{Fig:network}(a)). This framework can realize various quantum information-processing protocols, such as quantum repeater~\cite{briegel1998quantum}, quantum fingerprinting~\cite{arrazola2014quantum,Xufingerprinting}, blind quantum computing~\cite{dunjko2012blind}, and multiparty quantum communication~\cite{PhysRevLett.114.090501}. This scheme is advantageous in comparison to the recent demonstrations of quantum access networks~\cite{sasaki2011field,frohlich2013quantum,hughes2013network}, since it completely removes the need for the trust of the central relay node. Nevertheless, the scheme faces several crucial challenges in practice: (i) A key assumption is that the users' frequency-locked lasers are trusted. However, since frequency-locked lasers\footnote{See, for instance, http://dev.wavelengthreferences.com/clarity-precision-frequency-standard.} used in MDI-QKD experiments~\cite{Tittel:2012:MDI:exp, zhiyuan:experiment:2013,valivarthi2015measurement} are complicated apparatuses, there is a great risk involved in each user's trust that a commercial compact laser does not have any security loopholes. (ii) It is well known that the major challenge in implementation of Fig.~\ref{Fig:network}(a) is the performance of high-fidelity interference between photons from spatially separated lasers~\cite{Lo:MDIQKD,Tittel:2012:MDI:exp}. (iii) In fiber communication, it is necessary to introduce additional time-synchronization system and to include complex feedback controls to compensate for the polarization rotations (e.g., an implementation in~\cite{tang2014measurement}). All these challenges render Fig.~\ref{Fig:network}(a) difficult for practical implementations and applications.

Recently, to mitigate the experimental complexity of the interference from two remote lasers, several groups have proposed a new protocol against untrusted detectors~\cite{gonzalez2014quantum,lim2014detector, cao2014highly,liang2015simple}. A slight drawback is that a rigorous security analysis for this protocol is challenging, which makes the protocol vulnerable to attacks if certain assumptions cannot be satisfied~\cite{qi2014trustworthiness}. Another elegant proposal to resolve the limitations in the implementation of MDI-QKD is the so-called plug\&play MDI-QKD~\cite{kim2015plug}. Despite the importance of this proposal, a crucial part to guarantee the security -- source monitoring -- is ignored, which makes plug\&play MDI-QKD vulnerable to various source attacks~\cite{Gisin:attack:2006,Xu:phaseremapping:2010,sun2011passive}. Also, a complete security proof for plug\&play MDI-QKD and the analysis of practical imperfections are missing.

In this paper, we overcome the challenges of Fig.~\ref{Fig:network}(a) by proposing a MDI quantum network with a single untrusted source in Fig.~\ref{Fig:network}(b). The untrusted server transmits strong laser pulses to users, all of whom monitor the pulses, encode their bit information and send the attenuated pulses back to the server for measurement. We focus on the application of such a network to QKD. Crucially, we show that, even with an untrusted source, the communication security can be analyzed quantitatively and rigorously. Motivated by the security analysis for conventional plug\&play QKD \cite{zhao2008quantum,zhao2010security}, we show what measures by the users are necessary to ensure the security, and to rigorously derive a lower bound of the secure key generation rate. Moreover, we propose a novel decoy state method for MDI-QKD with an untrusted source. Furthermore, using simulations, we study how different real-life imperfections affect the security, and our simulation results show that MDI-QKD with an untrusted source provides a key generation rate that is close to the rate with trusted sources in the asymptotic limit. These results provide a complete security analysis for plug\&play MDI-QKD, and more importantly, make plug\&play MDI-QKD unconditionally secure, even with practical imperfections.

Our proposed MDI quantum network has the following advantages: (i) It completely removes the trust of the laser source. (ii) It can realize the MDI quantum network with a \emph{single} laser, which enables a high-fidelity interference among photons from different users. (iii) Due to the bi-directional structure, the system can automatically compensate for any birefringence effects and polarization-dependent losses in optical fibers, a feature that makes the system highly stable. (iv) The users can utilize the strong pulses from the server to easily synchronize and share time references. (v) There is a prospect of leveraging costly infrastructure for the quantum network, since the single laser source can be broadband, dynamically reconfigured and shared by several users via wavelength division multiplexing (WDM)~\cite{chapuran2009optical}.

The additional assumption, as compared to the initial MDI quantum network, is the trust of the monitoring devices. Note that the users need to monitor only classical laser pulses instead of single-photon signals. Such monitoring can be easily realized by a standard optical filter and a \emph{classical} intensity detector, and it is a necessary part of both BB84 and the initial MDI-QKD in order to prevent the so-called Trojan-horse attack~\cite{Gisin:attack:2006}. It is important that proof-of-concept experiments have been reported towards implementation of this monitoring~\cite{zhao2007experimental,peng2008experimental,sajeed2014attacks} and that ID Quantique's commercial system (i.e., Clavis2) has already included a preliminary version of the monitor~\cite{IDQ:company}. Recently, the security of the intensity detector has been studied comprehensively in~\cite{sajeed2014attacks}. Our work may lead to future research on an efficient implementation of the single-mode filtering and monitoring. This monitoring is also a key ingredient in other quantum communication protocols such as quantum illumination~\cite{zhang2013entanglement}.

The rest of this paper is organized as follows. We introduce the protocol of MDI-QKD with an untrusted source in Sec.~\ref{Sec:scheme}. In Sec.~\ref{Sec:analysis}, we present the security analysis of our protocol by introducing an equivalently virtual model. In Sec.~\ref{Sec:simulation}, we show the simulation results about the key rate comparison between MDI-QKD with an untrusted source and MDI-QKD with trusted sources, and study how device imperfections affect the protocol.  Finally, we conclude this paper in Sec.~\ref{Sec:discussion}.

\section{MDI-QKD with an untrusted source}~\label{Sec:scheme}

\begin{figure*}
\begin{center}
 \includegraphics[scale=0.85,angle=0]{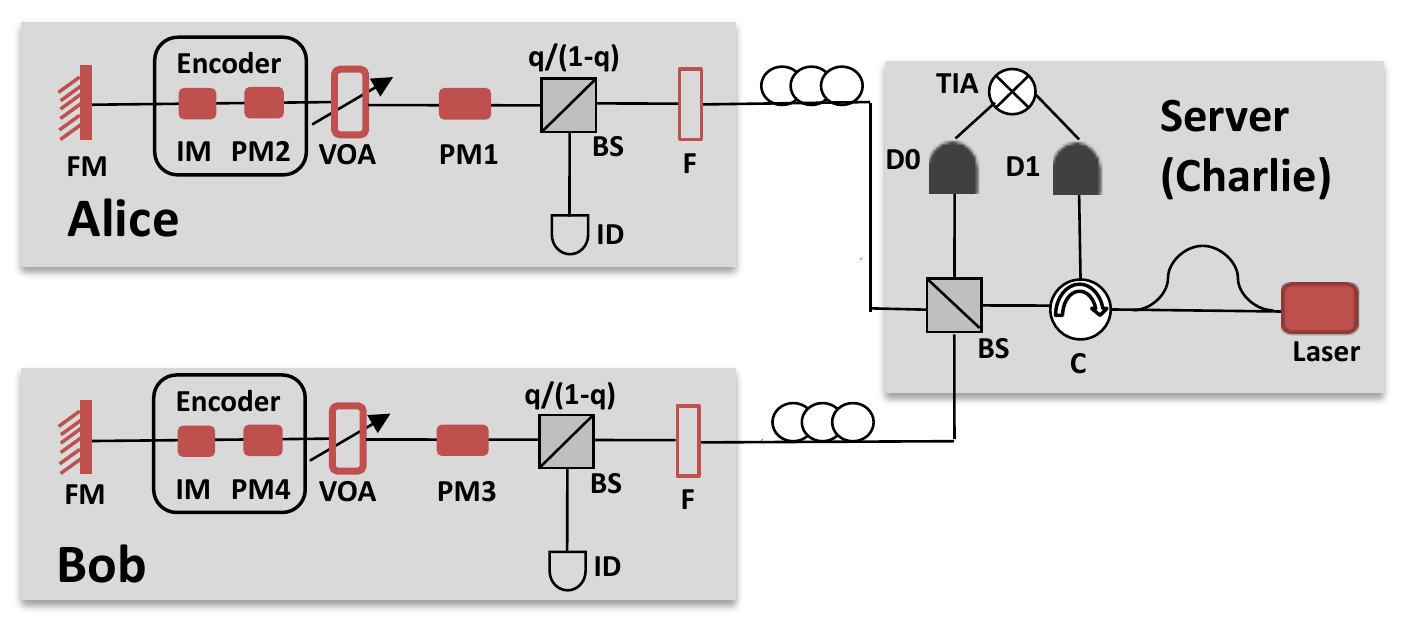}
 \end{center}
 \caption{(Color online) Schematic diagram of a time-bin-encoding MDI-QKD with an untrusted laser source. The strong time-bin laser pulses are generated by a pulsed laser and an interferometer in Charlie and they are split into two groups by a beam splitter (BS). Once these pulses arrived at Alice (Bob), they pass through an optical filter (F), a monitoring unit with a BS and a classical intensity detector (ID), a phase modulator (PM1 and PM3) for phase randomization and a variable optical attenuator (VOA). Then, the pulses are encoded by an Encoder that consists of an intensity modulator (IM) and a PM, and they are reflected by a Faraday mirror (FM). Finally, the pulses from Alice and Bob interfere at Charlie's BS and detected by two single photon detectors (D0 and D1). The coincident counts are recorded by a time interval analyzer (TIA). \label{Fig:setup}}
\end{figure*}

To illustrate our proposal, in Fig.~\ref{Fig:setup}, we present a specific design for QKD with two users. With simple modifications, our scheme can be applied to multiple users~\cite{PhysRevLett.114.090501}. We consider a time-bin encoding~\cite{Tittel:2012:MDI:exp}, and the protocol runs as follows.

\begin{description}
  \item[a. Preliminaries] Alice and Bob use a pre-shared key for authentication, and they negotiate parameters needed during the protocol run. Alice and Bob perform a calibration measurement of their devices.

  \item[b. Preparation and distribution] Charlie generates a strong laser pulse, which creates two time-bin pulses (early pulse and late pulse) after an interferometer. Charlie uses a beam splitter (BS) to split the two time-bin pulses into two parts and send them to Alice and Bob via two quantum channels (e.g., optical fibers).

  \item[c. Monitoring and Encoding phase] Once the pulses arrive at Alice (Bob), they pass through an optical filter, a monitoring unit, which consists of a BS and an intensity detector (ID). The pulses are phase-randomized by a phase modulator, PM1 (PM3), and then encoded by an Encoder that consists of an intensity modulator (IM) and a PM. Alice and Bob also use the IM to generate signal/decoy states. Finally, the pulses are reflected by a Faraday mirror (FM) and attenuated by a variable optical attenuator (VOA) to single-photon level.

  \item[c. Measurement phase] The time-bin-encoded weak coherent pulses from Alice and Bob travel back through the two channels, interfere at the BS of Charlie and finally they are detected by two single photon detectors. A coincident event projects the photons into the so-called singlet state $\ket{\psi^{-}}=(\ket{01}-\ket{10})/\sqrt{2}$~\cite{Tittel:2012:MDI:exp}.

  \item[d. Basis and signal/decoy reconciliation] Alice and Bob announce their encoding bases and signal/decoy intensity levels over the authenticated public channel and keep the samples measured in the same bases for signal/decoy states.

  \item[e. Parameter estimation] Alice and Bob perform the security analysis and the decoy state analysis based on their monitoring results and Charlie's public announcements.

  \item[g. Error reconciliation and privacy amplification] Alice and Bob perform the error correction. To ensure that they share a pair of identical keys,~they perform an error-verification step using two-universal hash functions. Finally, Alice and Bob apply the privacy amplification to produce the final secret key.
\end{description}

Since the source is entirely unknown and untrusted, we use three measures to enhance the security of our protocol \cite{Gisin:attack:2006,zhao2008quantum}.
\begin{enumerate}
  \item We place a narrow bandpass filter (together with a single mode fiber), i.e., F in Fig.~\ref{Fig:setup}, to allow only a single mode in spectral and spatial domains to enter into the Encoder. Note that a wavelength-bandpass filter has already been implemented in commercial QKD systems (i.e., Clavis2)~\cite{IDQ:company}. Moreover, the analysis in~\cite{xu2014experimental} shows that with standard optical devices, the single mode assumption can be guaranteed with a high rate of accuracy.

  \item We monitor the pulse energy and the arrival time to acquire certain information about the photon number distribution (PND) and the timing mode. Such monitoring can also defend the Trojan house attack~\cite{Gisin:attack:2006} and it has been included in commercial QKD systems~\cite{IDQ:company}. By randomly sampling the pulses to test the photon numbers, we can estimate some bounds on the output PND. In Fig.~\ref{Fig:setup}, this estimation is accomplished by Alice's and Bob's BS and ID. In practice, besides the ID, a spectrum analyzer can also be introduced to monitor the spectral information.

  \item Alice and Bob use PM1 and PM3 to apply the active phase randomization~\cite{zhao2007experimental}. The phase randomization is a general assumption made in most security proofs for laser-based QKD \cite{Hwang:2003, Lo:2005, Wang:2005} and the randomization can disentangle the input pulse into a classical mixture of Fock states.
\end{enumerate}

All the above three measures lead us to analyze the security of MDI-QKD with an untrusted source quantitatively and rigorously.

\section{Security analysis}~\label{Sec:analysis}

\subsection{System model}
\begin{figure*}
\begin{center}
 \includegraphics[scale=0.75,angle=0]{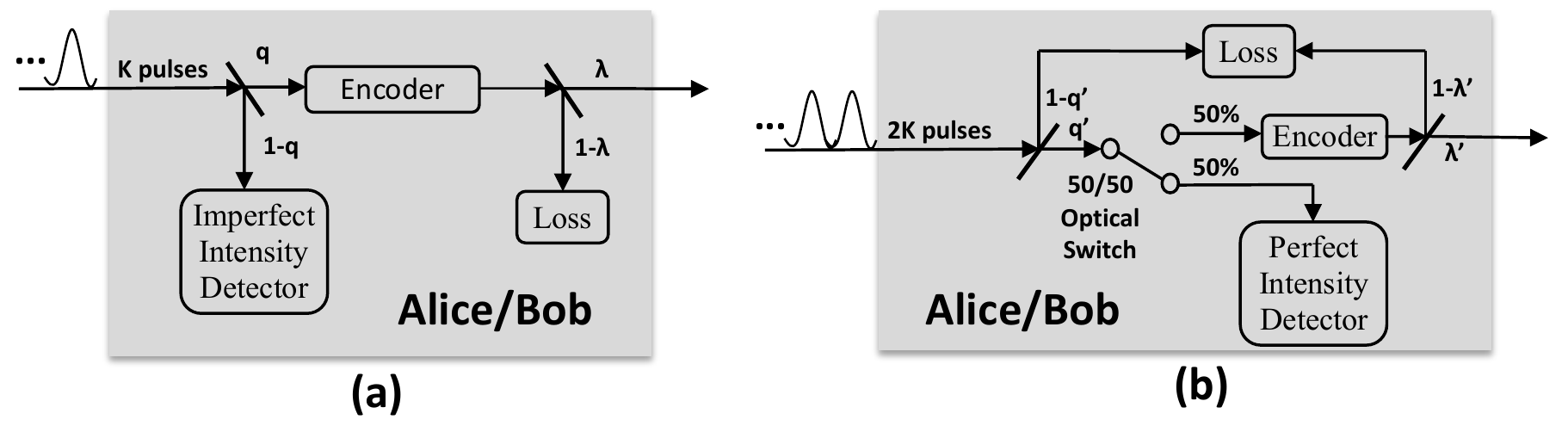}
 \end{center}
 \caption{(a) The actual model for Fig.~\ref{Fig:setup}. All the internal loss of Alice/Bob is modeled as a $\lambda/(1-\lambda)$ beam splitter. (b) An equivalent virtual model. The loss is modeled as a $\lambda'/(1-\lambda')$ beam splitter. $q'=\eta_\mathrm{ID}(1-q)$, where $\eta_\mathrm{ID}\le1$ is the efficiency of the imperfect intensity detector. $\lambda' = q\lambda/q'$. The virtual model, which has features from (a), is used to analyze the security of (a). \label{Fig:security}}
\end{figure*}

To analyze the security of Fig.~\ref{Fig:setup}, we model Alice's (Bob's) system in Fig.~\ref{Fig:security}(a). We model all the losses as a $\lambda/(1-\lambda)$ beam splitter, i.e., the internal transmittance of Alice's (Bob's) local lab is $\lambda_a$ ($\lambda_b$), which can be set accurately via VOA in Fig.~\ref{Fig:setup}. Each input pulse after the filter (F) is split into two via a BS: One (defined as the \emph{encoding pulse}) is sent to the Encoder for encoding, and the other (defined as the \emph{sampling pulse}) is sent to the ID for sampling. One might suppose that the PND of the encoding pulse could be easily estimated from the measurement result of the corresponding sampling pulse by using the random sampling theorem~\cite{papoulis2002probability,hoeffding1963probability}. However, this supposition is \emph{not} true. Any input pulse, after the phase randomization, is in a Fock state. Therefore, in the case of a pair of encoding and sampling pulses originating from the same input pulse, the PNDs of the two pulses are \emph{correlated}. This restriction suggests that the random sampling theorem cannot be directly applied.

We resolve the above restriction and analyze the security by introducing a virtual model in Fig.~\ref{Fig:security}(b) \cite{zhao2010security}. For the imperfect ID in Fig.~\ref{Fig:security}(a), assuming that its efficiency is $\eta_\mathrm{ID}\le 1$, we model the $q/(1-q)$ BS and the imperfect ID as a $q'/(1-q')$ BS and a perfect ID with $q'=(1-q)\eta_\mathrm{ID}$. To ensure that an identical attenuation is applied to the encoding pulses in both models, we redefine the internal transmittance in the virtual model as: $\lambda'=q\lambda/q'\le1$. Moreover, in the virtual model, we introduce a 50:50 optical switch to realize the active sampling. The optical switch, which is different from a BS, is solely a sampling device, without any restriction on the correlation of the PNDs of the encoding pulses and the sampling pulses. The random sampling theorem can be applied. A crucial fact is that the internal losses in the actual model and the virtual model are identical. The upper and lower bounds of output PND estimated from the virtual model are therefore also valid for those of the actual model, i.e., these two models are equivalent in the security analysis, an equivalence that has been proved in \cite{zhao2010security}.

In Fig.~\ref{Fig:security}(b), we define $m_a$ ($n_a$) as the photon number of the pulses that input (output) Alice. We also define the pulses that input Alice as
\begin{itemize}
  \item Untagged pulses:~~$m_a\in[(1-\delta_a)M_a,(1+\delta_a)M_a]$;
  \item Tagged pulses:~~$m_a<(1-\delta_a)M_a$ or $m_a>(1+\delta_a)M_a$.
\end{itemize}
Here $\delta_a$ is a small positive real number, and $M_a$ is a large positive integer (which can be the average of the input photon numbers of the pulses received by Alice). The same definitions apply to Bob's pulses with parameters $\{m_{b}, n_{b}, \delta_b, M_b\}$. Note that $\{\delta_a, \delta_b, M_a, M_b\}$ are chosen by Alice and Bob.


From the random sampling theorem, we draw the follow proposition~\cite{zhao2008quantum}.

\textbf{Proposition 1.} \emph{Consider that $2k$ pulses are sent to Alice from an untrusted source, and, of these pulse, $V_a$ pulses are untagged. Alice randomly assigns each pulse a status as either a sampling pulse or an encoding pulse with equal probabilities. In total, $V^\mathrm{s}_a$ sampling pulses and $V^\mathrm{e}_a$ encoding pulses are untagged. The probability that $V^\mathrm{e}_a\le V^\mathrm{s}_a-2\epsilon_a k$ satisfies
\begin{equation}\label{Eq:SamplingCodingDeviation}
\begin{aligned}
    P(V^\mathrm{e}_a\le V^\mathrm{s}_a-2\epsilon_a k)\le \exp(-k\epsilon_a^2),
\end{aligned}
\end{equation}
where $\epsilon_a$ is a small positive real number chosen by Alice (i.e. the error probability due to statistical fluctuations). That is, Alice can conclude that $V^\mathrm{e}_a>V^\mathrm{s}_a-\epsilon_a k$ with confidence level $\tau_a>1-\exp(-k\epsilon_a^2)$.}

The proof is shown in Appendix~\ref{App:Confidence_Active}. This proposition shows that Alice can estimate $V^\mathrm{e}_a$ from $V^\mathrm{s}_a$. Bob's untagged pulses have the same property. Furthermore, if we define $\Delta_a$ ($\Delta_b$) as the average probability that a sampling pulse belongs to a tagged sampling pulse in the asymptotic case, then Alice (Bob) can conclude that there are no fewer than $(1-\Delta_a-\epsilon_a)k$ ($(1-\Delta_b-\epsilon_b)k$) untagged encoding pulses with high fidelity.

\subsection{Untagged pulses}
In our analysis, Alice and Bob focus only on the untagged pulses for key generation and discard the other pulses. In practice, since Alice and Bob cannot perform quantum non-demolishing (QND) measurement on the photon number of the input pulses with current technology, they do not know which pulses are tagged and which are untagged. Consequently, the gain and the quantum bit error rate (QBER) of the untagged pulses can \emph{not} be measured directly. Alice and Bob can measure only the overall gain and the QBER. However, from Proposition 1, they know the probability that a certain pulse is tagged or untagged. Hence, they can estimate the upper and lower \emph{bounds} of the gain and the QBER of the untagged pulses. Furthermore, in the case that an untagged pulse inputs Alice or Bob in Fig.~\ref{Fig:security}(a), then the conditional probability that $n_a$ ($n_b$) photons are emitted by Alice obeys a binomial distribution, which can be controlled by Alice's (Bob's) internal transmittance. Alice (Bob) can also estimate the bounds of such a binomial distribution. The specific bounds for the gain, QBER and the PND of the untagged pulses are shown in Appendix~\ref{App:untagged}. Using these bounds, we can prove the security of MDI-QKD with an untrusted source quantitatively.

\subsection{Key rate} The secure key rate of MDI-QKD with an untrusted source in the asymptotic limit of infinite long keys is given by
\begin{equation} \label{Eqn:Key:formula}
\begin{aligned}
    R\geq (1-\Delta_a-\epsilon_a)(1-\Delta_b-\epsilon_b)\underline{Q_{11}^{\rm Z}}[1-H_{2}(\overline{e^{\rm X}_{11}})] \\
    -Q^{\rm Z}_{e, \mu\mu}f_{e}(E^{\rm Z}_{e, \mu\mu})H_{2}(E^{\rm Z}_{e, \mu\mu}),
\end{aligned}
\end{equation}
where $\underline{Q_{11}^{\rm Z}}$ and $\overline{e^{\rm X}_{11}}$ are, respectively, the lower bound of the gain in the rectilinear ($\rm Z$) basis and the upper bound of the error rate in the diagonal ($\rm X$) basis, given that both Alice and Bob send single-photon states in \emph{untagged} pulses; $H_{2}$ is the binary entropy function given by $H_2(x)$=$-x\log_2(x)-(1-x)\log_2(1-x)$; $Q^{\rm Z}_{e, \mu\mu}$ and $E^{\rm Z}_{e, \mu\mu}$ denote, respectively, the overall gain and QBER in the $\rm Z$ basis when Alice and Bob use signal states; $f_{e}\geq 1$ is the error correction inefficiency function (in simulations, we consider $f_{e}= 1.16$). Here we use the $Z$ basis for key generation and the $\rm X$ basis for testing only. In practice, $Q^{\rm Z}_{e, \mu\mu}$ and $E^{Z}_{e, \mu\mu}$ are directly measured in the experiment, while $\underline{Q_{11}^{\rm Z}}$ and $\overline{e^{\rm X}_{11}}$ are estimated from the decoy states.

\subsection{Decoy states}
In the previous decoy-state protocols for MDI-QKD~\cite{ma2012statistical, wang2013three, Feihu:practical, marcos:finite:2013, xu2014protocol}, the key assumption is that the yield, $Y_{n_an_b}$\footnote{$Y_{n_an_b}$ is defined as the conditional probability that Charlie has a coincident event given that Alice (Bob) sends out an $n_a$ ($n_b$) photon signal.}, remains the same for signal or decoy states. However, this assumption is \emph{no} longer valid in the case that the source is controlled by Eve, because Eve knows both the input photon number $m_a$ ($m_b$) and the output photon number $n_a$ ($n_b$). In this case, the parameter that is the \emph{same} for any signal and decoy state is $Y_{m_am_bn_an_b}$\footnote{$Y_{m_am_bn_an_b}$ is defined as the conditional probability that Charlie has a coincident event given that the two pulses enter Alice's and Bob's lab with photon number $m_a$ and $m_b$, and they emitted from Alice's and Bob's lab with photon number $n_a$ and $n_b$.}. Similarly, the conditional QBERs are also different if Eve controls the source.

Therefore, in MDI-QKD with an untrusted source, Eve is given significantly greater power, since she can control both the input and the output of Alice's and Bob's lab. The decoy state analysis is more challenging. However, rather surprisingly, it is still possible to achieve the unconditional security quantitatively, even if the source is given to Eve~\cite{zhao2008quantum}. This is so mainly because we are focusing only on the untagged pulses, whose PND, gain and QBER can be \emph{bounded}. Therefore, we are still able to estimate $\underline{Q^{\rm Z}_{11}}$ and $\overline{e^{\rm X}_{11}}$. Such an estimation can be completed by using either the numerical method based on linear programming or the analytical method. The details of this estimation are shown in Appendix~\ref{App:decoy}.

\section{Numerical simulation}~\label{Sec:simulation}
\begin{table}[hbt]
\centering
\begin{tabular}{c @{\hspace{0.3cm}} c @{\hspace{0.3cm}} c @{\hspace{0.3cm}} c @{\hspace{0.3cm}} c} \hline \hline
$\eta_{d}$ & $Y_{0}$ & $e_{d}$ & f & $\alpha$ \\
20\% & $3\times 10^{-6}$ & 0.1\% & 75 MHz & 0.21 dB/km \\
\hline
$\eta_{ID}$ & $\sigma_{ID}$ & $q$ & $\epsilon$ & $k$ \\
0.7 & $6.55\times10^{4}$ & 0.01 & $10^{-10}$ & $3.5\times10^{13}$ \\
\hline \hline
\end{tabular}
\caption{List of practical parameters for simulation. The detection efficiency $\eta_{d}$ and the dark count rate $Y_{0}$ are from commercial ID-220 single photon detectors \cite{IDQ:company}. The channel misalignment error $e_{d}$, the system repetition rate $f$, the total number of pulses $k$ and the fiber loss coefficient $\alpha$ are from the 200 km MDI-QKD experiment~\cite{tang2014measurement}. The efficiency of the intensity detector (ID) $\eta_{ID}$, the noise of the ID $\sigma_{ID}$, and the beam splitter ratio $q$ are from~\cite{zhao2010security}. $\epsilon$ is the security bound considered in our finite-key analysis.} \label{Tab:exp:parameters}
\end{table}

The details of the simulation techniques, including the model for the imperfect intensity detector, the tagged ratio $\Delta$ and the finite-data statistics, are shown in the Appendix~\ref{App:simulation}. We use the experimental parameters, listed in Table~\ref{Tab:exp:parameters} for simulation. For $\delta=\delta_a=\delta_b$, a choice for it that is too large or too small will make the security analysis less optimal~\cite{zhao2010security}. We find numerically that $\delta=0.01$ is a near an optimal value. In addition, we assume that the source in Charlie is Poissonian centered at $M_\mathrm{c}$ photons per optical pulse. In this bi-directional structure, Alice's and Bob's average input photon numbers ($M_a$ and $M_b$) depend on the channel loss and $M_\mathrm{c}$. The gain and the QBER are derived using the channel model presented in~\cite{Feihu:practical}.

\begin{figure}
  \includegraphics[width=8cm]{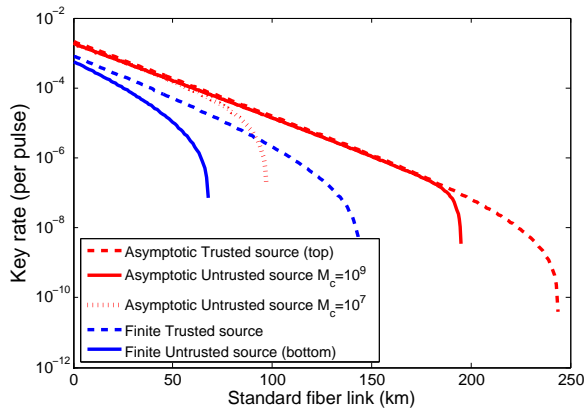}\\
  \caption{(Color online) Simulation results. Red curves are for an infinite number of signals. $M_\mathrm{c}$ denotes the mean of the photon number (per pulse) of Charlie's laser pulses. With $M_\mathrm{c}=10^9$ and practical imperfections, MDI-QKD with an untrsuted source can tolerate about 195 km distance. At the distances below 180 km, the key rates for the two cases (with trusted and untrusted source) are almost overlapping. Blue curves are for a finite number of signals. Finite data size reduces the efficiencies for both cases. MDI-QKD with an untrusted source can still tolerate over 70 km fiber with detectors of 20\% efficiency. }\label{FIG:infinite:ImpID}
\end{figure}

The simulation results with an infinite number of signals are shown by the red curves in Fig.~\ref{FIG:infinite:ImpID}. We consider two decoy states: we fix the vacuum state at $\omega=0$, the weak decoy state at $\nu=0.01$, and optimize the signal state $\mu$ for different distances. With $M_\mathrm{c}=10^{7}$ (Charlie's mean photon number per pulse), the case with an untrusted source (red dotted curve) is similar to that with trusted sources (red dashed curve) at short distances. The condition changes at long distances. This occurs because at long distances, due to the channel loss, the photon numbers arrived at by Alice and Bob will be much smaller than $M_c$. The lower input photon number increases $\Delta$ and the estimate of the gain of the untagged pulses is sensitive to the value of $\Delta$ (see Eq.~\eqref{eq:Qbound}). This is so especially when the measured overall gain is small over long distances. In contrast, over short distances, the gain is significantly greater than $\Delta$; therefore, the key rates for the two cases are almost overlapping.

A natural scheme for the improvement of the performance of MDI-QKD with an untrusted source is the use of a brighter laser. Indeed, the performance is improved substantially by setting $M_\mathrm{c}=10^9$ (red solid curve). The two cases (with trusted sources and with an untrusted source) have similar results. Note that sub-nanosecond pulses with $\sim10^9$ photons per pulse can be easily generated with directly modulated laser diodes. For instance, if the wavelength is 1550 nm and the pulse repetition rate is 75 MHz, the average laser power of Charlie's source is $\sim9.6$ mW. This laser power can be provided by many commercial pulsed laser diodes.


The simulation results with a finite number of signals are shown by the blue curves in Fig.~\ref{FIG:infinite:ImpID}. We choose the confidence level $\tau$ for the statistical fluctuations of the estimation of the number of untagged pulses (see Eq. \eqref{Eq:SamplingCodingDeviation}) as $\tau_a=\tau_b=\tau \ge
1-10^{-7}$, which suggests that $\epsilon_a=\epsilon_b=3.03\times10^{-7}$. We set $M_\mathrm{c}=10^9$. We can see that finite data size clearly reduces the efficiencies: first, the statistical fluctuation for decoy-state MDI-QKD becomes important, and this factor reduces the performance of both the trusted source and the untrusted source. Second, $\epsilon_a$ and $\epsilon_b$ are non-zero in this finite data case, and thus the estimate of the gain of the untagged pulses becomes not tight (see Appendix~\ref{App:simulation})\footnote{That is, due to statistical fluctuations, the proportion of tagged pulses is increased. Our analysis is conservative in that Eve can fully control the tagged pulses, which makes the security bound worse than MDI-QKD with trusted sources.}. In the finite data setting, our protocol can tolerate about 70 km fiber with standard commercial detectors of 20\% efficiency. With state-of-the-art detectors~\cite{marsili2013detecting}, however, the protocol can easily generate keys over 200 km fiber.

\section{Discussion}~\label{Sec:discussion}
In summary, we propose a MDI quantum network with an untrusted source. In this network, the complicated and expensive detectors, together with the laser source, can be provided by an untrusted network server that can be shared by all users; that is, a star-type MDI quantum access network can be readily realized on the basis of our proposal for several quantum information processing protocols~\cite{xu2014measurement,briegel1998quantum,arrazola2014quantum,Xufingerprinting,dunjko2012blind}. Our work proves the feasibility of such a realization. Moreover, we present a complete security analysis for MDI-QKD with an untrusted source. Our analysis and simulation consider various practical imperfections, including additional loss introduced by the bi-directional structure, the inefficiency and noise of the intensity monitor, and the finite data-size effect. Furthermore, our protocol is practically secure and ready for implementation. An experimental demonstration is in progress.

It is worthy to mention that one practical issue associated with Fig.~\ref{Fig:setup} is the temporal matching. For instance, the two channels -- Alice-Charlie and Bob-Charlie -- may be different in practice, then the arrival times are mismatch when the pulses return back to Charlie. One solution, as demonstrated already in~\cite{Tittel:2012:MDI:exp}, is to introduce addition fibers inside either Alice or Bob to match the channel length. Note that, similar to the conventional plug\&play system~\cite{IDQ:company}, in our proposal, each of Alice and Bob should hold certain length of fiber spool as the optical buffer. Consequently, Alice/Bob can control the length of this fiber buffer to match the channel difference. Furthermore, the feedback control techniques, demonstrated in~\cite{tang2014measurement} could also be used to ameliorate temporal-matching issues.

There are still several imperfections that are not analyzed in our paper, such as the source flaws in the state preparation~\cite{xu2014experimental}, the non-ideality in the optical filter, and the imperfections in the electronics of the classical intensity detector~\cite{sajeed2014attacks}. These questions lead to an interesting future project, that is, deriving a refined analysis that includes all possible (small) imperfections and side channels in users. For instance, Ref.~\cite{lucamarini2015practical} has reported comprehensive analysis against the Trojan horse attack. We expect that this research direction will move an important step towards unconditionally secure communication networks that are also practical.

\section{Acknowledgement}
We thank H.-K. Lo, B. Qi and S.~Sun for valuable discussions. Support from the Office of Naval Research (ONR) and the Air Force Office of Scientific Research (AFOSR) is gratefully acknowledged.

\appendix

\section{Proof of Proposition 1}~\label{App:Confidence_Active}
We follow \cite{zhao2010security} to prove Proposition 1. Among all the $V$ untagged pulses, each pulse has probability 1/2 to be assigned as an untagged encoding pulse. Therefore, the probability that $V^\mathrm{e}_a=v$ obeys a binomial distribution. Cumulative probability is given by \cite{hoeffding1963probability}
\begin{equation*}
    P(V^\mathrm{e}_a\le\frac{V-2\epsilon k}{2}|V=v) \le \exp(-\frac{4\epsilon^2k^2}{v})
\end{equation*}

For any $v\in[0,2k]$, $2k/v\ge 1$. Therefore, we have
\begin{equation*}
    P(V^\mathrm{e}_a\le\frac{V-2\epsilon k}{2}|V\in[0, 2k]) \le \exp(-k\epsilon^2).
\end{equation*}

Since $V\in[0,2k]$ is always true, the above inequality reduces to
\begin{equation}\label{Eq:CumulativeBinomial}
    P(V^\mathrm{e}_a\le\frac{V-2\epsilon k}{2}) \le \exp(-k\epsilon^2).
\end{equation}

By definition, we have
\begin{equation}\label{Eq:SumUntagged}
V=V^\mathrm{e}_a+V^\mathrm{s}_a.
\end{equation}
Substituting Equation \eqref{Eq:SumUntagged} into Equation
\eqref{Eq:CumulativeBinomial}, we have
\begin{equation}\label{Eq:SamplingCodingDeviation_Proof}
    P(V^\mathrm{e}_a\le V^\mathrm{s}_a-\epsilon k)\le \exp(-k\epsilon^2).\qed
\end{equation}

The above proof can be easily generalized to the case where for each pulse sent from the untrusted source to Alice/Bob, Alice/Bob randomly assigns it as either an encoding pulse with probability $\beta$, or a sampling pulse with probability $1-\beta$. Here $\beta\in(0,1)$ is chosen by Alice/Bob. It is then straightforward to show that
\begin{equation}\label{Eq:SamplingCodingDeviation_General}
    P[V^\mathrm{e}_a\le \frac{\beta}{1-\beta}(V^\mathrm{s}_a-2\epsilon k)] \le \exp(-4k\epsilon^2\beta^2).
\end{equation}

\section{Properties of untagged pulses}~\label{App:untagged}
The main concept to analyze the properties of the untagged pulses follows the analysis for plug\&play QKD presented in~\cite{zhao2008quantum}. Both Alice and Bob will focus on the $(1-\Delta_a-\epsilon_a)k$ and $(1-\Delta_b-\epsilon_b)k$ untagged pulses for key generation and discard the other pulses. This provides a conservative way to analyze the security, and also, owing to the input photon numbers of the untagged pulses concentrated within a narrow range, this makes it much easier to analyze the security.

In practice, since Alice and Bob cannot perform quantum non-demolishing measurement on the photon number of the input pulses with current technology, they do not know which pulses are tagged and which are untagged. As a result, the gain $Q$ and the quantum bit error rate (QBER) $E$ of the untagged pules cannot be measured experimentally. Here $Q$ is defined as the \emph{conditional}
probability that Charlie has a coincident event given that both Alice and Bob send out an untagged pulse and Alice and Bob use the same basis; $E$ is defined as error rates inside $Q$.

In experiment, Alice and Bob can measure the overall gain $Q_e$ and the overall QBER $E_e$. The subscript $e$ denotes the experimentally measurable overall properties. Moreover, they know the probability that certain pulse to be tagged or untagged from the above analysis. Although they cannot measure the gain $Q$ and the QBER $E$ of the untagged pulses directly, they can estimate the upper bounds and lower bounds of them. The upper bound and lower bound of $Q$ are:
\begin{equation}\label{eq:Qbound}
\begin{aligned}
    Q\leq \overline{Q} &= \frac{Q_e}{(1-\Delta_a-\epsilon_a)(1-\Delta_b-\epsilon_b)},\\
    Q\geq \underline{Q} &= \max(0, \frac{Q_e-1+(1-\Delta_a-\epsilon_a)(1-\Delta_b-\epsilon_b)}{(1-\Delta_a-\epsilon_a)(1-\Delta_b-\epsilon_b)}).
\end{aligned}
\end{equation}
The upper bound and lower bound of $E\cdot Q$ can be estimated as
\begin{equation}\label{eq:Ebound}
    \begin{aligned}
    \overline{E\cdot Q} &=
    \frac{Q_e E_e}{(1-\Delta_a-\epsilon_a)(1-\Delta_b-\epsilon_b)},\\
    \underline{E\cdot Q} &=
    \max(0,\frac{Q_e E_e-1+(1-\Delta_a-\epsilon_a)(1-\Delta_b-\epsilon_b)}{(1-\Delta_a-\epsilon_a)(1-\Delta_b-\epsilon_b)}).
    \end{aligned}
\end{equation}

Moreover, suppose that an untagged pulse with input photon number $m_a\in[(1-\delta_a)M_a,(1+\delta_a)M_a]$ inputs Fig.3(a) of main-text, the conditional probability that $n_a$ photons are emitted by Alice given that $m_a$ photons enter Alice obeys binomial distribution as:
\begin{align}\label{eq:pn}
    P(n_a|m_a) &= {m_a\choose n_a} (\lambda_aq)^{n_a}(1-\lambda_aq)^{m_a-n_a} & (0\le\lambda_a\le1)
\end{align}

For Alice's untagged bits, we can show that the upper bound and lower bound of $P(n_a|m_a)$ are:
\begin{widetext}
\begin{equation}\label{eq:pnbound}
\begin{aligned}
    \overline{P(n_a|m_a)} &= \left\{
                        \begin{array}{ll}
                         (1-\lambda_aq)^{(1-\delta_a)M_a}, & \hbox{if $n_a=0$;}\\
                          {(1+\delta_a)M_a\choose n_a}(\lambda_aq)^{n_a}(1-\lambda_aq)^{(1+\delta_a)M_a-n_a}, & \hbox{if $1\le n \le (1+\delta_a)M_a$;} \\
                          0, & \hbox{if $n_a>(1+\delta_a)M_a$;}
                        \end{array}
                      \right.\\
    \underline{P(n_a|m_a)} &= \left\{
                        \begin{array}{ll}
                         (1-\lambda_aq)^{(1+\delta_a)M_a}, & \hbox{if $n_a=0$;}\\
                          {(1-\delta_a)M_a\choose n_a}(\lambda_aq)^{n_a}(1-\lambda_aq)^{(1-\delta_a)M_a-n_a}, & \hbox{if $1\le n \le (1-\delta_a)M_a$;} \\
                          0, & \hbox{if $n_a>(1-\delta_a)M_a$;}
                        \end{array}
                      \right.
\end{aligned}
\end{equation}
\end{widetext}
under the condition: $(1+\delta_a)M_a\lambda_aq<1$. This condition suggests that the expected output photon number of any
untagged pulse should be lower than 1. This is normally a basic condition in decoy-state BB84 and MDI-QKD based on weak coherent pulses. For example, for $M_a=10^7$ and $q=0.01$, Alice can simply set $\lambda_a=10^{-6}$ so that the expected output photon number is 0.1.

\section{Decoy state analysis} \label{App:decoy}
Various decoy-state methods have been proposed for MDI-QKD~\cite{ma2012statistical,wang2013three,xu2014protocol}. Among all these decoy state protocols, the two decoy state protocol has been shown to be the optimal one~\cite{xu2014protocol}, it has already been used in all experimental MDI-QKD implementations reported so far~\cite{Tittel:2012:MDI:exp, Liu:2012:MDI:exp, da2012proof, zhiyuan:experiment:2013, tang2014measurement, valivarthi2015measurement}. In this protocol, there are three states: Alice's signal state $\mu_a$ (for which the internal transmittance is $\lambda^{\mu}_a$), Alice's two weak decoy states $\nu_a$ and $\omega_a$ (for which the internal transmittance is $\lambda^{\omega}_a<\lambda^{\nu}_a<\lambda^{\mu}_a$). In this work, we focus on the \emph{symmetric} case where the two channel transmissions from Alice to Charlie and from Bob to Charlie are equal. In symmetric case, the optimal intensities for Alice and Bob are equal~\cite{xu2014protocol}. Hence, to simplify our discussion, we assume that equal intensities are used by Alice and Bob, \ie, $\gamma_a$=$\gamma_b$=$\gamma$ with $\gamma\in\{\mu,\nu,\omega\}$. Also, we consider that only the signal state is used to generate the final key, while the decoy states are solely used to test the channel properties.

In previous decoy-state protocols for MDI-QKD~\cite{ma2012statistical,wang2013three,xu2014protocol}, the key assumption is that the yield of $n_a$ and $n_b$ photon state $Y_{n_an_b}$ remains the same, whatever signal states or decoy states are chosen by Alice and Bob, e.g. $Y_{n_an_b}^{\mu\mu}=Y_{n_an_b}^{\nu\nu}$. Here $Y_{n_an_b}^{\mu\mu}$ is defined as the conditional probability that Charlie has a coincident event given that Alice (Bob) sends out an $n_a$ ($n_b$) photon signal and they both chose signal state by setting internal transmittances $\lambda_a^{\mu}$ and $\lambda_b^{\mu}$. This is true because in previous analysis, Eve knows only the output photon numbers $n_a$ and $n_b$ of each pulse. However, this assumption is \emph{no} longer valid in the case that the source is controlled by Eve. Because Eve knows both the input photon number $m_a$ ($m_b$) and the output photon number $n_a$ ($n_b$) when she controls the source. Therefore she can perform an attack that depends on the values of both $m$ and $n$. In this case, the parameter that is the \emph{same} for any signal and decoy states is $Y_{m_am_bn_an_b}$, the conditional probability that Charlie has a coincident event given that the two pulses enter Alice's and Bob's lab with photon number $m_a$ and $m_b$, and they emitted from Alice's and Bob's lab with photon number $n_a$ and $n_b$. Similarly, the conditional QBERs are also different: $e_{n_an_b}^{\mu\mu}\neq e_{n_an_b}^{\nu\nu}$ if Eve controls the source. The parameter that is the same for the signal state and the decoy states is $e_{m_am_bn_an_b}$.

In summary, in MDI-QKD, if the source is assumed to be trusted, we have:
\begin{equation} \nonumber
\begin{aligned}
& Y_{n_an_b}^{\mu\mu}=Y_{n_an_b}^{\nu\nu} \\
& e_{n_an_b}^{\mu\mu}=e_{n_an_b}^{\nu\nu}.
\end{aligned}
\end{equation}
If the source is accessible to Eve (i.e., the source is untrusted), we have:
\begin{equation} \nonumber
\begin{aligned}
& Y_{m_am_bn_an_b}^{\mu\mu}=Y_{m_am_bn_an_b}^{\nu\nu} \\
& e_{m_am_bn_an_b}^{\mu\mu}=e_{m_am_bn_an_b}^{\nu\nu}.
\end{aligned}
\end{equation}
The dependence of $Y_{n_an_b}$ and $e_{n_an_b}$ on different states is a fundamental difference between MDI-QKD with an untrusted source and MDI-QKD with trusted source. Therefore, in MDI-QKD with an untrusted source, Eve is given significantly greater power, and the decoy state analysis is much more \emph{challenging}. However, rather surprisingly, it is still possible to achieve the unconditional security quantitatively even if the source is given to Eve. This is mainly because we are only focusing on the untagged pulses, whose photon number distribution, the gain and the QBER can be \emph{bounded} via Eqs. \eqref{eq:pnbound}, \eqref{eq:Qbound}, \eqref{eq:Ebound} respectively. Therefore we are still able to estimate $\underline{Q^{\rm Z}_{11}}$ and $\overline{e^{\rm X}_{11}}$. Such estimation can be completed by using either numerical method based on linear programming or analytical method.

In a MDI-QKD implementation with an untrusted source, by performing the measurements for different intensity settings, we can obtain:
\begin{widetext}
\begin{equation} \label{Eqn:general:decoy}
\begin{aligned}
&Q_{\gamma_a\gamma_b}^{\chi}=\sum_{m_a=(1-\delta_a)M_a}^{m_a=(1+\delta_a)M_a}\sum_{m_b=(1-\delta_b)M_b}^{m_b=(1+\delta_b)M_b}\sum_{n_a=0}^{\infty}
\sum_{n_b=0}^{\infty}P_{in}(m_a)P_{in}(m_b)P^{\gamma_a}(n_a|m_a)P^{\gamma_b}(n_b|m_b)Y_{m_am_bn_an_b} \\
&E_{\gamma_a\gamma_b}^{\chi}Q_{\gamma_a\gamma_b}^{\chi}=\sum_{m_a=(1-\delta_a)M_a}^{m_a=(1+\delta_a)M_a}\sum_{m_b=(1-\delta_b)M_b}^{m_b=(1+\delta_b)M_b}\sum_{n_a=0}^{\infty}
\sum_{n_b=0}^{\infty}P_{in}(m_a)P_{in}(m_b)P^{\gamma_a}(n_a|m_a)P^{\gamma_b}(n_b|m_b)Y_{m_am_bn_an_b}e_{m_am_bn_an_b}
\end{aligned}
\end{equation}
\end{widetext}
where $\chi\in\{\rm X,\rm Z\}$ denotes the basis choice, $\gamma_a$ ($\gamma_b$) denotes Alice's (Bob's) intensity setting, $Q_{\gamma_a\gamma_b}^{\chi}$ ($E_{\gamma_a\gamma_b}^{\chi}$) denotes the gain (QBER); where $P_\text{in}(m_a)$ is the probability that the input signal contains $m_a$ photons (i.e., the ratio of the number of signals with $m$ input photons over $k$), $P^{\gamma_a}(n_a|m_a)$ is the conditional probability that the output signal contains $n_a$ photons given the
input signal contains $m_a$ photons, for state $\gamma_a$ and is given by Eq. \eqref{eq:pn}.

$Q^{\rm Z}_{11}$ for $\gamma_a=\mu$ and $\gamma_b=\mu$ can be written as
\begin{widetext}
\begin{equation}  \label{Eq:S11}
\begin{aligned}
Q^{\rm Z}_{11} & =\sum_{m_a=(1-\delta_a)M_a}^{m_a=(1+\delta_a)M_a}\sum_{m_b=(1-\delta_b)M_b}^{m_b=(1+\delta_b)M_b}
P_{in}(m_a)P_{in}(m_b)P^{\mu}(1|m_a)P^{\mu}(1|m_b)Y_{m_am_b11} \\
& \geq \underline{P^{\mu}_{1|m_a}}\underline{P^{\mu}_{1|m_b}} \sum_{m_a=(1-\delta_a)M_a}^{m_a=(1+\delta_a)M_a}\sum_{m_b=(1-\delta_b)M_b}^{m_b=(1+\delta_b)M_b}
P_{in}(m_a)P_{in}(m_b)Y_{m_am_b11} \equiv \underline{P^{\mu}_{1|m_a}}\underline{P^{\mu}_{1|m_b}} S^{\rm Z}_{11},
\end{aligned}
\end{equation}
\end{widetext}
where the bounds of the probabilities are from Eqs. \eqref{eq:pnbound}. Thus, the estimation on $\underline{Q^{\rm Z}_{11}}$ is equivalent to the estimation of $\underline{S^{\rm Z}_{11}}$, and Eq.~(\ref{Eqn:general:decoy}) can be written as
\begin{equation} \label{Eqn:general:decoysimplify}
\begin{aligned}
& Q_{\gamma_a\gamma_b}^{\chi}=\sum_{n_a, n_b=0}^{\infty}P^{\gamma_a}(n_a|m_a)P^{\gamma_b}(n_b|m_b)S^{\chi}_{n_an_b} \\
& E_{\gamma_a\gamma_b}^{\chi}Q_{\gamma_a\gamma_b}^{\chi}=\sum_{n_a, n_b=0}^{\infty}P^{\gamma_a}(n_a|m_a)P^{\gamma_b}(n_b|m_b)S^{\chi}_{n_an_b}e^{\chi}_{n_an_b}
\end{aligned}
\end{equation}

\subsection{Numerical approaches} \label{App:numerical}
Ignoring statistical fluctuations temporally, the estimations on $\underline{S^{\rm Z}_{11}}$ and $\overline{e^{\rm X}_{11}}$, from Eq.~(\ref{Eqn:general:decoysimplify}) are constrained optimisation problems, which is linear and can be efficiently solved by linear programming (LP). The numerical routine to solve these problems can be written as:
\begin{widetext}
\begin{equation} \label{Eqn:decoy:numerical} \nonumber
\begin{aligned}
min: \space & S^{\rm Z}_{11}, \\
s.t.: \space & 0\leq S_{n_an_b}^{\rm Z}\leq 1, with \ n_a,n_b\in {\mathcal S}_{\rm cut}; \\
\overline{P(n_a|m_a)} &= \left\{
                        \begin{array}{ll}
                         (1-\lambda_a)^{(1-\delta_a)M_a}, & \hbox{if $n_a=0$;}\\
                          {(1+\delta_a)M_a\choose n_a}\lambda^{n_a}(1-\lambda_a)^{(1+\delta_a)M_a-n_a}, & \hbox{if $1\le n \le (1+\delta_a)M_a$;} \\
                          0, & \hbox{if $n_a>(1+\delta_a)M_a$;}
                        \end{array}
                      \right.\\
    \underline{P(n_a|m_a)} &= \left\{
                        \begin{array}{ll}
                         (1-\lambda_a)^{(1+\delta_a)M_a}, & \hbox{if $n_a=0$;}\\
                          {(1-\delta_a)M_a\choose n_a}\lambda^{n_a}(1-\lambda_a)^{(1-\delta_a)M_a-n_a}, & \hbox{if $1\le n \le (1-\delta_a)M_a$;} \\
                          0, & \hbox{if $n_a>(1-\delta_a)M_a$;}
                        \end{array}
                      \right. \\
& \underline{Q_{\gamma_a\gamma_b}^{\rm Z}}-1+\sum_{n_a,n_b\in S_{cut}}\underline{P^{\gamma_a}(n_a|m_a)}\underline{P^{\gamma_b}(n_b|m_b)} \leq \sum_{n,m\in {\mathcal S}_{\rm cut}}P^{\gamma_a}(n_a|m_a)P^{\gamma_b}(n_b|m_b)S^{\rm Z}_{n_an_b}\leq \overline{Q_{\gamma_a\gamma_b}^{\rm Z}} \\ \\
Max: \space &e_{11}^{\rm X}, \\
s.t.: \space & 0\leq S_{n_an_b}^{\rm X}\leq 1, 0\leq S_{n_an_b}^{\rm X}e_{n_an_b}^{\rm X}\leq 1, with \ n_a,n_b\in {\mathcal S}_{\rm cut} \\
\overline{P(n_a|m_a)} &= \left\{
                        \begin{array}{ll}
                         (1-\lambda_aq)^{(1-\delta_a)M_a}, & \hbox{if $n_a=0$;}\\
                          {(1+\delta_a)M_a\choose n_a}(\lambda_aq)^{n_a}(1-\lambda_aq)^{(1+\delta_a)M_a-n_a}, & \hbox{if $1\le n \le (1+\delta_a)M_a$;} \\
                          0, & \hbox{if $n_a>(1+\delta_a)M_a$;}
                        \end{array}
                      \right.\\
    \underline{P(n_a|m_a)} &= \left\{
                        \begin{array}{ll}
                         (1-\lambda_aq)^{(1+\delta_a)M_a}, & \hbox{if $n_a=0$;}\\
                          {(1-\delta_a)M_a\choose n_a}(\lambda_aq)^{n_a}(1-\lambda_aq)^{(1-\delta_a)M_a-n_a}, & \hbox{if $1\le n \le (1-\delta_a)M_a$;} \\
                          0, & \hbox{if $n_a>(1-\delta_a)M_a$;}
                        \end{array}
                      \right. \\
& \underline{Q_{\gamma_a\gamma_b}^{\rm X}}-1+\sum_{n_a,n_b\in S_{cut}}\underline{P^{\gamma_a}(n_a|m_a)}\underline{P^{\gamma_b}(n_b|m_b)} \leq \sum_{n,m\in {\mathcal S}_{\rm cut}}P^{\gamma_a}(n_a|m_a)P^{\gamma_b}(n_b|m_b)S^{\rm X}_{n_an_b} \leq \overline{Q_{\gamma_a\gamma_b}^{\rm X}} \\
\underline{Q_{\gamma_a\gamma_b}^{\rm X}E_{\gamma_a\gamma_b}^{\rm X}} & -1+\sum_{n_a,n_b\in S_{cut}}\underline{P^{\gamma_a}(n_a|m_a)}\underline{P^{\gamma_b}(n_b|m_b)} \leq \sum_{n,m\in {\mathcal S}_{\rm cut}}P^{\gamma_a}(n_a|m_a)P^{\gamma_b}(n_b|m_b)S^{\rm X}_{n_an_b}e^{\rm X}_{n_an_b} \leq \overline{Q_{\gamma_a\gamma_b}^{\rm X}E_{\gamma_a\gamma_b}^{\rm X}},
\end{aligned}
\end{equation}
\end{widetext}
where ${\mathcal S}_{\rm cut}$ denotes a finite set of indexes $n_a$ and $n_b$, with ${\mathcal S}_{\rm cut}=\{n_a,n_b \in {\mathbb N} \ {\rm with}\ n_a\leq{}A_{\rm cut}$ and $n_b\leq{}B_{\rm cut} \}$,
for prefixed values of $A_{\rm cut}\geq{}2$ and $NB_{\rm cut}\geq{}2$. In our simulations, we choose $A_{\rm cut}=7$ and $B_{\rm cut}=7$, as larger $A_{\rm cut}$ and $B_{\rm cut}$ have negligible effect on decoy-state estimation. More discussions can be seen in~\cite{ma2012statistical}. Here, $\gamma\in\{\mu,\nu,\omega\}$ for two decoy-state estimation. Notice that statistical fluctuations can be easily conducted by adding constraints on the experimental measurements of $Q_{\gamma_a\gamma_b}^{\chi}$ and $E_{\gamma_a\gamma_b}^{\chi}$. These additional constraints can be analyzed by using statistical estimation methods, such as standard error analysis~\cite{ma2012statistical} or Chernoff bound~\cite{marcos:finite:2013}. A rigorous finite-key analysis can also be implemented by following the technique presented in~\cite{marcos:finite:2013}.
\subsection{Analytical approaches} \label{Sec:analytical}
A rigorous estimation is to solve the equation set of Eq.~(\ref{Eqn:general:decoysimplify}) by using the constrains on the binomial probability distributions given by Eq.~(\ref{eq:pnbound}). The analytical expression for such an estimation is highly complicated. So, we only use numerical method presented in last section to study this precise estimation. Here, for the analytical expression, we present a relatively simple analytical method by using the Poisson limit theorem~\cite{papoulis2002probability}:

\textbf{Claim:} Under the condition that $m\rightarrow \infty$ and $\lambda q \rightarrow 0$, such that $\mu=m\lambda q$, then
\begin{align}\label{eq:approximation}
{m\choose n} (\lambda q)^{n}(1-\lambda q)^{m-n} \rightarrow \exp(-\mu)\frac{\mu^n}{n!}
\end{align}

The condition in this claim is easy to meet in an actual experiment as $m$ can be larger than $10^{6}$ and $\lambda q$ is normally lower that $10^{-7}$ in a practical setup. The intuition behind this approximation is that we applied heavy attenuation on the input pulses in Alice and bob. The input pulse has more than $\sim10^6$ photons, while the output pulse has less than one photon on average. The internal attenuation of Alice's local lab is greater than -60dB. We know that heavy attenuation will transform arbitrary photon number distribution into a Poisson-like distribution. A qualitative argument on this argument for the plug-and-play structure has been provided in \cite{Gisin:attack:2006}. From the approximation, Eq.~(\ref{Eqn:general:decoysimplify}) can be estimated using the similar methods presented in \cite{xu2014protocol}.

The lower bound of $S^{\rm Z}_{11}$ is given by
\begin{widetext}
\begin{equation} \nonumber
\begin{aligned}
\underline{S^{\rm Z}_{11}}
& =\frac{1}{(\mu-\omega)^2(\nu-\omega)^2(\mu-\nu)}\times[(\mu^2-\omega^2)(\mu-\omega)(\underline{Q_{\nu\nu}^{\rm Z}}e^{2\nu}+\underline{Q_{\omega\omega}^{\rm Z}}e^{2\omega}-\overline{Q_{\nu\omega}^{\rm Z}}e^{\nu+\omega}-\overline{Q_{\omega\nu}^{\rm Z}}e^{\omega+\nu})-\\
& (\nu^2-\omega^2)(\nu-\omega)(\overline{Q_{\mu\mu}^{\rm Z}}e^{2\mu}+ \overline{Q_{\omega\omega}^{\rm Z}}e^{2\omega}-\underline{Q_{\mu\omega}^{\rm Z}}e^{\mu+\omega}-\underline{Q_{\omega\mu}^{\rm Z}}e^{\omega+\mu})].
\end{aligned}
\end{equation}
\end{widetext}

The upper bound of $S^{\rm X}_{11}e^{\rm X}_{11}$ is given by
\begin{widetext}
\begin{equation} \nonumber
\begin{aligned}
\overline{S^{\rm X}_{11}e^{\rm X}_{11}}=\frac{1}{(\nu-\omega)^2}\times [e^{2\nu}\overline{Q_{\nu\nu}^{\rm X}E_{\nu\nu}^{\rm X}}+e^{2\omega}\overline{Q_{\omega\omega}^{\rm X}E_{\omega\omega}^{\rm X}}-e^{\nu+\omega}\underline{Q_{\nu\omega}^{\rm X}E_{\nu\omega}^{\rm X}}-e^{\omega+\nu}\underline{Q_{\omega\nu}^{\rm X}E_{\omega\nu}^{\rm X}}].
\end{aligned}
\end{equation}
\end{widetext}

\section{Simulation techniques}~\label{App:simulation}
In simulation, the gain and the QBER are derived using the channel model presented in~\cite{Feihu:practical}. We consider two decoy states: $\nu=0.01$ and $\omega=0$, and we optimize the signal state $\mu$ for different distances. We choose $f_{e}= 1.16$

\subsection{Imperfect intensity detector}
There are two major imperfections of the intensity detector (ID): inefficiency and noise. The inefficiency $\eta_{ID}$ can be easily modeled as additional loss by using a beam splitter. The noise of the ID is another important imperfection. In a real experiment, the ID may indicate a certain pulse contains $m'$ photons. Here we refer to $m'$ as the \emph{measured} photon number in contrast to the actual photon number $m$. However, due to the noise and the inaccuracy of the intensity monitor, this pulse may not contain exactly $m'$ photons. To quantify this imperfection, following~\cite{zhao2010security}, we introduce a term, called conservative interval $\varsigma$. We then define $\underline{V}^\mathrm{s}$ as the number of sampling pulses with measured photon number $m'\in[(1-\delta)M'+\varsigma,(1+\delta)M'-\varsigma]$, where $M'=M\eta_{ID}(1-q)$. One can conclude that, with confidence level $\tau_\mathrm{c}=1-c(\varsigma)$, the number of untagged sampling pulses $V^\mathrm{s}\ge \underline{V}^\mathrm{s}$. One can make $c(\varsigma)$ arbitrarily close to 0 by choosing a large enough $\varsigma$. That is, for one individual pulse, the probability that $|m-m'|>\varsigma$ can be negligible.

In practice, various noise sources, including thermal noise, shot-noise, etc, may exist. Here, in simulation, we consider a simple noise model where a constant Gaussian noise with variance $\sigma_\mathrm{ID}^2$ is assumed. That is, if $m$ photons enter an efficient but noisy ID, the probability that the measured photon number is $m'$ obeys a Gaussian distribution
\begin{equation}\label{eq:Pmm}
P(m'|m)=\frac{1}{\sigma_\mathrm{IM}\sqrt{2\pi}}\exp[-\frac{(m-m')^2}{2\sigma_\mathrm{ID}^2}].
\end{equation}
Hence, the measured photon number distribution $P(m')$ has a larger variation than the actual photon number distribution $P(m)$ due to the noise. More concretely, if the input photon numbers obeys a Gaussian distribution centered at $M$ with variance
$\sigma^2$, the measured photon numbers also obeys a Gaussian distribution centered at $M'$, but with a variance $\sigma^2+\sigma_\mathrm{ID}^2$.

\subsection{The tagged ratio $\Delta$}
For any $\delta\in[0,1]$ and the imperfect ID discussed above, we can calculate $\Delta$ from the measured photon number $m'$ by
\begin{equation}\label{eq:deltap}
\Delta = 1-[\Phi(M'+\delta M'+\varsigma)-\Phi(M'-\delta M'-\varsigma))],
\end{equation}
where $\Phi$ is the cumulative distribution function of the photon number for the measured pulses~\cite{papoulis2002probability}. Assuming that the system is based on a coherent source by Charlie, which means that the input photon number $m$ obeys Poisson distribution. It is natural to set $M$ to be the average input photon number. In numerical simulation, for ease of calculation, we approximate the Poisson distribution of the input photon number $M$ as a Gaussian distribution centered at $M$ with variance $\sigma^2=M$. This is an excellent approximation because $M$ is very large ($10^7$ or larger) in all the simulations presented below. Then, the measured photon number $m'$ follows a Gaussian distribution centered at $M'=M\eta_{ID}(1-q)$ with a variance $M+\sigma_\mathrm{ID}^2$. The Gaussian cumulative distribution function is given by~\cite{papoulis2002probability}
\begin{equation}\label{eq:cumulative_gaussian}
    \Phi_g(x) = \frac{1}{2}[1+\text{erf}(\frac{x-M'}{\sqrt{2(M+\sigma_\mathrm{ID}^2)}})],
\end{equation}
where $\text{erf}(x) = \frac{2}{\sqrt{\pi}}\int_0^xe^{-t^2}dt$ is the error function. Notice that erf$(x)$ is an odd function, from Eqs. \eqref{eq:deltap} and \eqref{eq:cumulative_gaussian}, we have
\begin{equation}\label{eq:deltap_gaussian}
    \begin{aligned}
        \Delta = 1-\text{erf}(\frac{\delta M'+\varsigma}{\sqrt{2M+2\sigma_\mathrm{ID}^2}}).
    \end{aligned}
\end{equation}

In simulation, for $\delta=\delta_a=\delta_b$, a choice for it that is too large or too small will make the security analysis less optimal~\cite{zhao2010security}. We find numerically that $\delta=0.01$ is a near an optimal value.

\subsection{Finite-data statistics}
A real-life QKD experiment is always completed in finite time, which means that the length of the output secret key is
obviously finite. Thus, the parameter estimation procedure in QKD needs to take the statistical fluctuations of the different parameters into account. We assume that Charlie's source generates $2k$ pulses in an experiment. The finite data effect has two main consequences: First, the finite data size will introduce statistical fluctuations for the estimation of the number of untagged pulses. If the confidence level $\tau_a$ for Proposition 1 is expected to be close to 1, $\epsilon_a$ has to be positive. More concretely, for a fixed $2k$, if the estimate on the untrusted source is expected to have confidence level no less than $\tau_a$, one has to choose $\epsilon_a$ as $\epsilon_\mathrm{a} = \sqrt{-\frac{\ln(1-\tau_a)}{k}}$. In simulation, we choose the confidence level $\tau$ (see Proposition 1) as $\tau_a=\tau_b=\tau \ge 1-10^{-7}$, which suggests that $\epsilon_a=\epsilon_b=3.03\times10^{-7}$. Since $\epsilon_a$ and $\epsilon_b$ are non-zero in this finite data case, the estimate of the gain of the untagged pulses becomes not tight at long distances. That is, due to statistical fluctuations, the proportion of tagged pulses is increased at long distance. Our analysis is \emph{conservative} in that Eve can fully control the tagged pulses, which makes the security bounds worse than MDI-QKD with trusted sources. This is the reason why MDI-QKD with an untrusted source is not as good as MDI-QKD with trusted sources in the finite-data case, which has been shown in the Fig.~\ref{FIG:infinite:ImpID} of main text.

Second, in decoy state MDI-QKD, the statistical fluctuations of experimental outputs have to be considered. The technique to analyze the statistical fluctuations can be analyzed by using statistical estimation methods, such as standard error analysis~\cite{ma2012statistical} or Chernoff bound~\cite{marcos:finite:2013}. In this paper, we analyze the statistical fluctuations by using the standard error analysis method presented in~\cite{ma2012statistical}. In simulation, we choose $\epsilon=10^{-10}$ as the overall security bound.

\bibliographystyle{ieeetr}
\bibliography{MDIuntrust}

\end{document}